\documentclass[a4paper,showkeys,showpacs,floatfix,aps,pre,reprint,groupedaddress]{revtex4-1}

\usepackage{graphicx}
\usepackage{amsmath, amssymb, amsfonts}
\usepackage{color}
\usepackage{subfig}
\usepackage{caption}
\usepackage{comment}

\newcommand{\dd}[1]{\ensuremath{\,\mathrm{d}#1}}

\begin{document}

\title{Effect of fractal disorder on static friction in the Tomlinson
model}



\author{Jon Alm Eriksen$^{1,2}$}
\email[]{jon.alm.eriksen@gmail.com}
\author{Soumyajyoti Biswas$^{2}$}
\email[]{soumyajyoti.biswas@saha.ac.in}
\author{Bikas K. Chakrabarti$^{2}$}
\email[]{bikask.chakrabarti@saha.ac.in}

\affiliation{
$^{1}$
Department of Physics, Norwegian University of Science and Technology,
N-7491 Trondheim, Norway.\\
$^{2}$
Theoretical Condensed Matter Physics Division, Saha Institute of Nuclear 
Physics, 1/AF Bidhannagar, Kolkata-700064, India.\\
}

\date{August 28, 2010}

\begin{abstract}
\noindent We propose a modified version of the Tomlinson model for
static friction
between two chains of beads. We introduce disorder in terms of vacancies in the
chain, and distribute the remaining beads in a scale invariant way. For this we utilize a generalized random Cantor set. We
relate the static friction force, to the overlap distribution of the
chains, and discuss how the distribution of the static friction
force depends on the distribution of the remaining beads. For the
random Cantor set we find a scaled distribution which is independent on the generation of the set.
\end{abstract}


\maketitle

\section{Introduction}

\noindent Friction is a long studied phenomenon in physics \cite{bowden,bnj,braun}. Still, the regime of validity and the microscopic origin of the 
empirical laws of static and dynamic friction (that goes under the
name of Amontons-Coulomb) are not well established.  But the
advancements of technology in the last few decades has triggered both
theoretical \cite{gnecco,sang,kajita,capozza,kawaguchi,caroli2}, as well as
experimental investigations in this field \cite{mate, hirano}.  

Surfaces which appear smooth may contain roughness at the micrometer
scale \cite{caroli} as well as impurities in the contact region
\cite{muser} which will
influence the friction properties of the materials.
If we limit ourselves to the study of atomically clean 
surfaces,  it is sufficient  to
consider the relative motion of the atomic layers in contact.
The earliest attempt to model such a situation was carried out by Tomlinson
\cite{tomlinson}. Tomlinson's model is a chain of
beads, representing atoms, all of which are individually attached to a
body above. The chain is dragged on a periodic potential representing a corrugated
substrate (see Fig. \ref{model} (a)). Each bead interacts with the
potential, but there is no interaction among the beads.
As in the Tomlinson model for dry friction (see e.g. \cite{weiss}), the surface atoms
are considered  to be independent oscillators capable of
absorbing a finite energy  and momentum.
Although they are connected by the chain,  the energy is assumed to
get dissipated within the bulk and not transferred  to the
neighbouring  atoms along the chain.
This model has regained its importance in recent years
because, in experiments using atomic force microscopy, the contact
probe can be studied at an atomic level (see e.g., \cite{braun}). There are also a number of similar models using chains of beads
moving over a substrate. Most notably is the Frenkel-Kontorova model \cite{fk1},
where the beads also interact with the nearest neighbors.

In this paper, we propose a modified version of the Tomlinson
model. We consider a two chain version of the model,
where one of the chains slides on top of the other (see
Fig. \ref{model}
(b)). We study the effect of disorder in the static friction
force. In particular, we introduce vacancy disorder, by removing
beads, in both chains. We relate the
static friction force to the measure of overlap of beads. The disorder
in the chains is such that the remaining beads are distributed in a
scale invariant way. For this we use a generalized version of the
Cantor set. This leads to a certain kind of overlap distribution, and
we compare this to other ways of distributing the beads, and discuss
how this relates to the distribution of static friction force.

It may be mentioned at this point that rough surfaces have a
self-affine character and therefore can be represented by self-similar
fractals \cite{mandelbrot}. 
We represent the disorder in the Tomlinson model by a generalized
random Cantor set, which is one of the simplest examples of a
fractal. Although the Cantor set is a purely mathematical
construction, it has been studied extensively by physicist as it
encapsulates the essential property of self-similarity and scale invariance.

The paper is organized in the following way: In section (II) we define
the model. Then in section (III) we briefly discuss the basic
definitions and useful properties of the Cantor set, which will be
required in the subsequent discussions. In section (IV) we find the
expressions for the static friction force distribution for 
Tomlinson's model with beads
distributed as random cantor sets. We conclude by
summarizing the results in section (V).

\section{The Model}
\noindent 

The static friction force $F_s$, is defined to be the
minimum force needed to initiate sliding between two objects in
contact. If the applied force exceeds this threshold $F>F_s$, the
objects will move relative to each other. The presence of this force
is due to a local minimum in the energy landscape of spatial configurations of the two
objects. Two disordered surfaces may however have several
configurations which correspond to local
minima, and the static friction may have different values
in the different configurations.

\begin{figure}[htb]
  \begin{center}
    \includegraphics[scale=0.35]{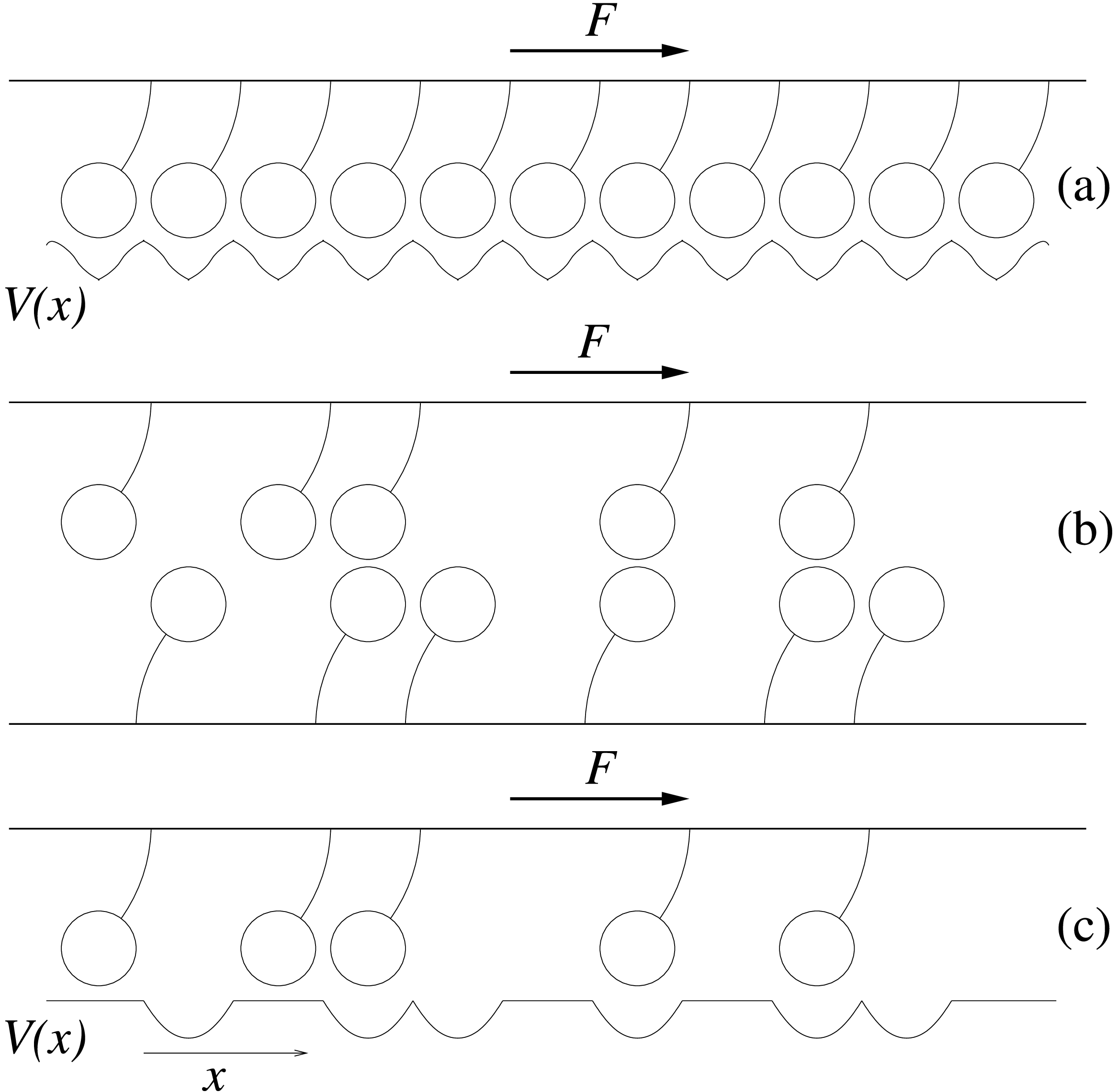}
   \caption{
(a) The Tomlinson model: A chain of beads being
  dragged with an applied force $F$ in a periodic potential
  $V(x)$. (b) a two chain version of the Tomlinson model, with defects
  in terms of vacancies. (c) Same as (b) with the potential from the
  underlying chain illustrated.
}
\label{model}
\end{center}
\end{figure}

We consider a two chain version of Tomlinson's model, where one chain slides on top of the
other (see Fig. \ref{model} (b)). We introduce defects in terms of
vacancies (removed beads).
The vacancies are introduced randomly, and are statistically identical in both chains.
The spacings between the sites in the chain (either occupied by a bead or removed), are
constant and equal in both chains. 

The underlying chain gives rise to a substrate potential for the chain above. 
We assume that the interaction potential between the beads in the opposite chains are
short ranged and attractive, such that the substrate potential is on
the form of a series of potential wells, corresponding to the
remaining beads of the chain below (see Fig. \ref{model} (c)).

The potential energy of a configuration of this
modified Tomlinson model will be determined by the sum of beads locked in potential wells. 
The static friction force of one bead locked in a
potential well is determined by the maximum value of the derivative of
the potential. We will not specify the functional form of the
potential well, but we assume that the static friction force of one
bead in a potential has a given value. The static friction force of
a chain will therefore be directly proportional to the number of beads
locked in potential wells.

If there were no disorder, the overlap would be constant and the static friction force per bead would
also be constant. If, however, there is disorder in terms of vacancies,
the number of locked beads (referred to as the overlap)
will not be constant but will have a set of possible values. These values, and hence
the static friction force, will follow a distribution.

We assume that the expected number of overlapping beads is
given and investigate how the distribution vary around the
expectation value.
We will see that this distribution will have certain features if the
beads (and the potential wells from the substrate) are distributed with a
spatial correlation between them. This distribution will differ
significantly from the case when the remaining beads (and the
potential wells from the substrate) are distributed uniformly.

\begin{figure}[htb]
  \begin{center}
    \includegraphics[scale=0.48]{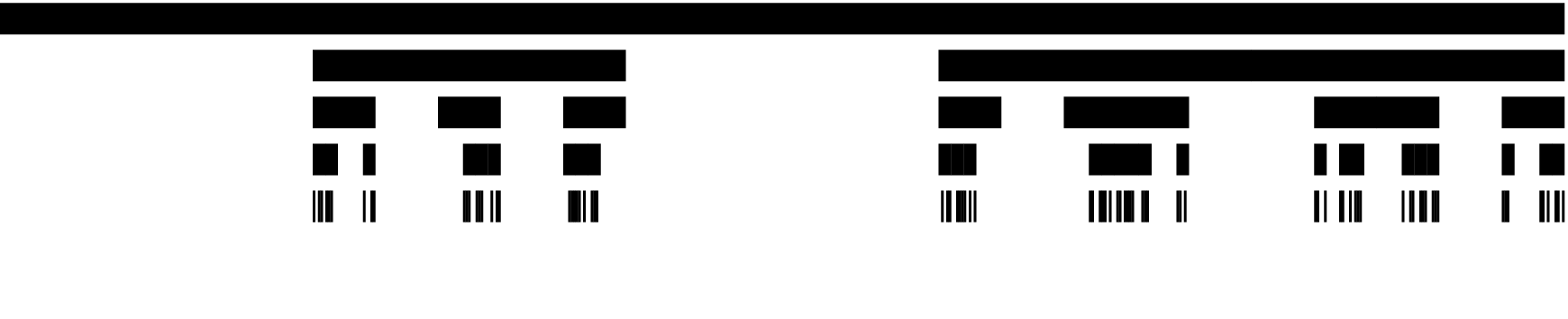}
   \caption{Construction of a random Cantor set, $s=5$ and $r=3$.}
   \label{rCantore}
  \end{center}
\end{figure}

We are especially interested in the properties of the distribution when
the beads are distributed with a spatial correlation
which decays as a power-law in the distance between the beads, as
this correspond to a fractal disorder in the chain.
A binary chain (with vacant or occupied sites) is, however, not uniquely
determined by the spatial correlation structure. We choose to utilize
generalized random Cantor sets for the distribution of the remaining beads (see
Fig. \ref{rCantore}), as the random Cantor set has the desired correlation
structure on an average (this is a result of the self-similar property
of the set). Moreover the structure of the random Cantor set is simple enough to allow
theoretical results for the overlap distribution.

\section{Cantor sets}
\label{cantorsec}

\noindent The prototype example of a fractal is the Cantor set
$C$. The Cantor set is constructed by first removing the middle third
of the base interval [0, 1]. From each of the remaining intervals, [0,
1/3] and [2/3, 1], a middle third is again removed. The process of removing middle thirds of the remaining intervals is continued ad infinitum. The intervals which are left after the middle third of every remaining interval have been removed $n$ times, is referred to as the Cantor set $C_n$, at generation $n$. $C_n$ becomes a true fractal as $n$ goes to infinity $\displaystyle C=\lim_{n \rightarrow \infty} C_{n}$.

There are two self-similar transformations related to the Cantor set
$C$. The transformations are $S^{(1)}(x)=x/3$ and
$S^{(2)}(x)=(x+2)/3$. We can describe the process of removing
the middle third of every interval by the action of these
transformation,

\begin{equation*}
C_n = S^{(1)}(C_{n-1})  \bigcup S^{(2)}(C_{n-1}). 
\end{equation*}

\begin{figure}[htb]
  \begin{center}
    \includegraphics[scale=0.48]{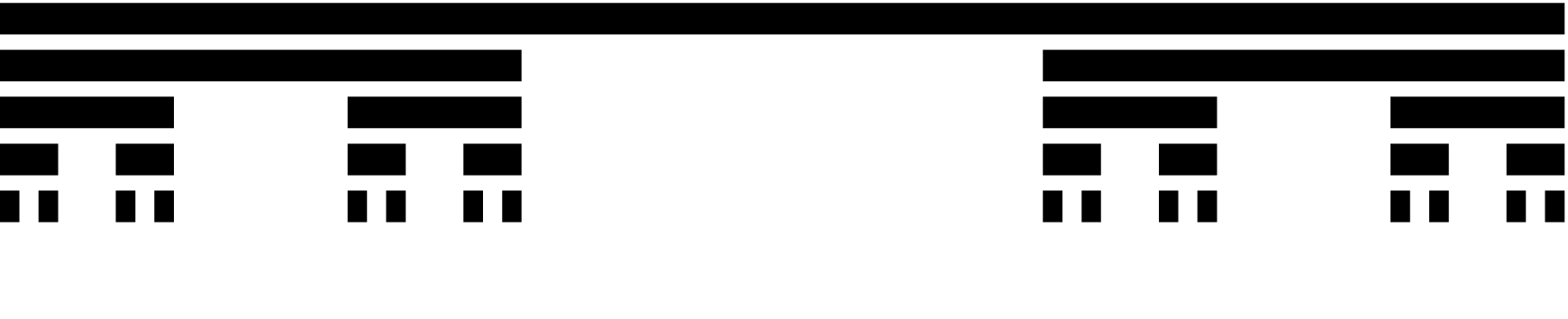}
   \caption{Construction of a triadic Cantor set.}
   \label{triCant}
  \end{center}
\end{figure}

\noindent And we can define $C$ to be the subset of $[0,1]$ which is invariant
under the union of these two transformations. Fig. \ref{triCant}
illustrates the construction procedure.

In order to investigate properties which emerge from the fractal
nature of a subset of an interval, we will extend the notion of a
Cantor set. We will construct a generalized Cantor set by the action
of $r$ self-similar transformations on the interval [0, 1]. Let
$a=\{a_1,a_2,...,a_r\}$ be a set of integers such that $1 < r < s$ and $a_i<a_j$ for $i<j$. The self-similar transformations be on the following form:

\begin{equation*}
S^{(i)}(x) = \frac{x+a_i-1}{s}.
\end{equation*}

\noindent And we will obtain the generalized Cantor set in the obvious way

\begin{equation*}
C_n = \bigcup_i S^{(i)}(C_{n-1}).
\end{equation*}

The Cantor set constructed this way will consist of $r^n$ line
elements of equal length that may or may not be connected. We identify
the integers in $a$ as the positions of the remaining line elements at
the first generation. The Box Counting dimension of this generalized
Cantor set is given by $D_B=\log(r)/\log(s)$ (see e.g. \cite{falconer}),  and the regular triadic Cantor set is obtained by taking $s=3$ and $a=\{1,3\}$. 

The notion of a random Cantor set is ambiguous as we can randomize 
in several ways. We will define the random version of this generalized
Cantor set analogous to Falconer's definition of the triadic Cantor set
in \cite{falconer}. For each generational
step, divide every remaining line element from the previous generation
into $s$ elements of equal length and remove all but $r$ of them. For
each line segment, and for each generation, randomize the position of
the $r$ remaining intervals, but keep the number of remaining
intervals fixed. The self-similar properties of this set hold only in
a statistical sense. 

The generalized Cantor set will at generation $n$ consist of $r^n$ of
the $s^n$ equal distant parts of the interval [0,1]. This enables us
to represent it as a vector $(C_{n,1}, C_{n,2}, ...,C_{n,s^n})$. The
vector element $C_{n,i}$ represent the $i$th interval of the Cantor
set such that $C_{n,i}$ takes the value 1 if the interval
$[(i-1)/s,i/s]$ is contained in $C_n$, and the value 0 if it is
removed. The vector contains therefore $s^n$ elements, out of which
$r^n$ takes the value 1, and the rest takes the value 0.

The Cantor set is of measure zero. That is, the size of the Cantor set
(sum of the remaining line segments), will go to zero as the generation
goes to infinity. For that reason we study the Cantor set at a finite generation, at the atomic level there
is, of course, no way to continue the removal procedure.

\section{Overlap}
\subsection{Overlap of random sets}

\noindent Let us first consider the overlap of two chains with
randomly placed vacancies, uniformly distributed along the chain. The
overlap distribution is trivial, but we include it
here as it will be instructive to compare it with the other results.

Consider two independently generated vectors $\left(X_1,...,X_N \right)$,
$\left(X'_1,...,X'_N\right)$ containing $N$ binary variates
representing the chain (0 represents a removed bead at a given site). 
Each element takes the value 1 with probability $p$. The probability
of an overlap at element $i$ is
therefore $\Pr(X_i=1,X'_i=1)=p^{2} $. The overlap of the two vectors is given by $Y =
\sum X_i X'_i$. And the overlap distribution is on the form of a binomial distribution:

\begin{align}
\Pr(Y=x)={N \choose x} p^{2x} (1-p^2)^{N-x}.
\label{randoverlap}
\end{align}

\subsection{Overlap of random Cantor sets}

\begin{figure}[htb]
\begin{center}
  \includegraphics[scale=0.48]{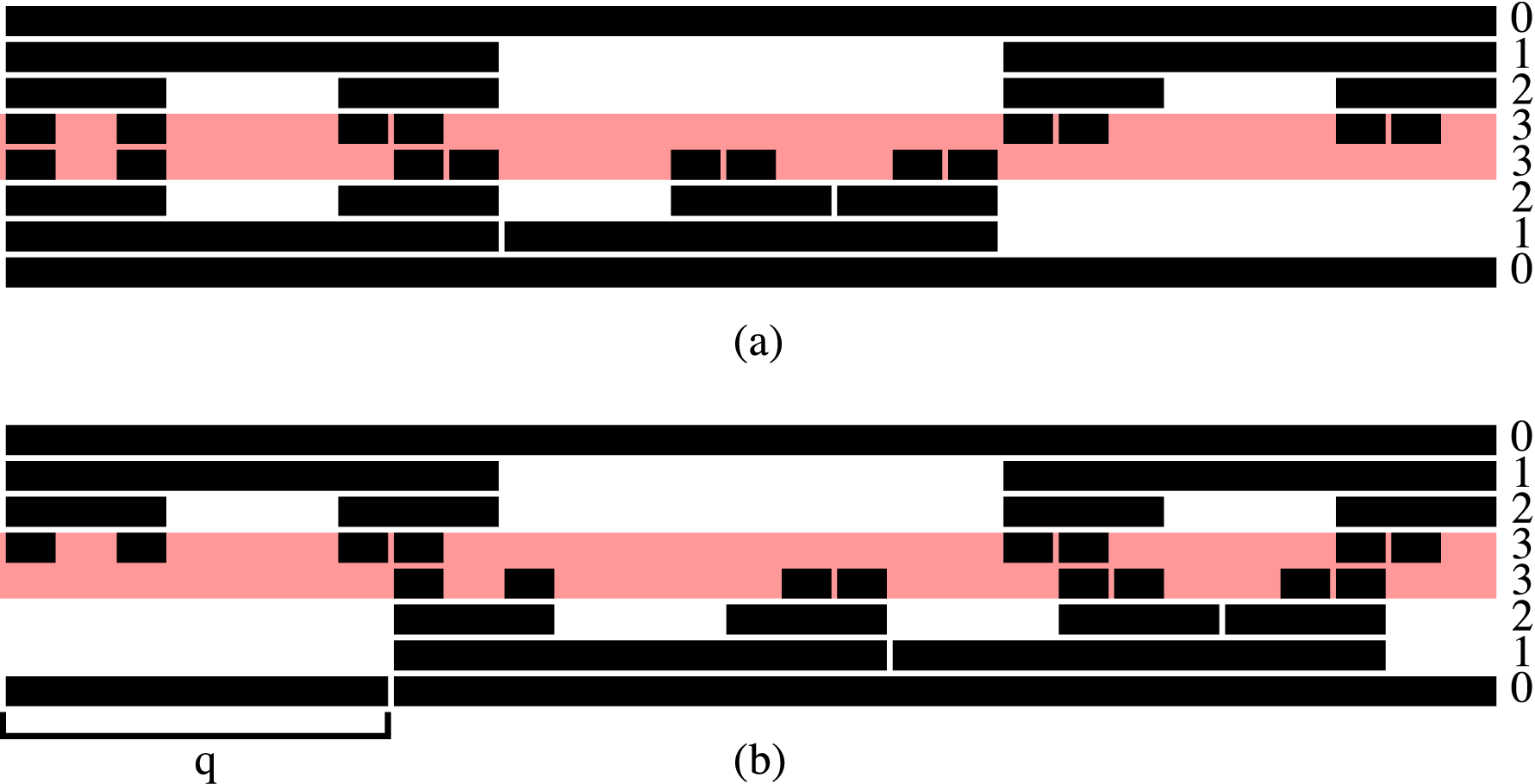}
  \caption{(Color online)
The overlap of two (independently generated) random Cantor sets at
    generation 3 ($s=3, r=2$). The generation number of the two
    sets are marked on the right side. (a) shows the overlap when the
    sets are perfectly aligned, and (b) shows the overlap when the
    sets are shifted relatively by an offset $q$. The overlap size is the number of overlapping elements at the final
    generation of the opposite sets (in the shaded region). The number of overlapping elements are 3 for both (a) and (b).}
\label{model1}
\end{center}
\end{figure}

\noindent In order to describe the overlap distribution for two random
(independently generated)
Cantor sets, $C_{n,i}$ and $C'_{n,i}$, it is instructive to consider first the overlap of
two random triadic Cantor sets ($s=3, r=2$), without a relative
displacement as illustrated in Fig. \ref{model1} (a). 
Notice, however, that Fig. \ref{model1} also illustrates the \emph{construction} of the Cantor
sets. The figure does not suggest that the Cantor set is embedded in
two dimensions. The overlap at generation
$n$ is given by $Y_n=\sum C_{n,i} C'_{n,i} $. For $n=1$ one can easily calculate, by summing up the possible outcomes, that the probability distribution of the overlap takes the following form

\begin{align*}
\Pr(Y_1=0) &= 0,\\
\Pr(Y_1=1) &= 2/3,\\
\Pr(Y_1=2) &= 1/3.\\
\end{align*}

Call $A$ the event that only one element overlap at the first generation
of the construction, and $B$ the event that two elements overlap at the
first generation of the construction. $A$ and $B$ are mutually exclusive events, and $P(A)+P(B)=1$. Consider the
case when $(C_{n,i})$ is constructed by an iterative procedure like
in Fig. \ref{rCantore}.  We can write the following relation

\begin{equation*}
\Pr(Y_n=x) = \frac{2}{3}\Pr(Y_n=x|A)+\frac{1}{3}\Pr(Y_n=x|B).
\end{equation*}

\noindent Note that

\begin{align*}
\Pr(Y_{n}=x|A)&=\Pr(Y_{n-1}=x), \\
\Pr(Y_{n}=x|B)&=\sum_{i \leq x} \Pr(Y_{n-1}=x-i)\Pr(Y_{n-1}=i). 
\end{align*}

\noindent So we have

\begin{align*}
&\quad \Pr(Y_n=x)=\\
&\frac{2}{3} \Pr(Y_{n-1}=x)+\frac{1}{3}\sum_{i\leq x} \Pr(Y_{n-1}=x-i) \Pr(Y_{n-1}=i).
\end{align*}

As $n$ grows large, we can approximate this by a continuous
distribution $f_n(x)=\Pr(Y_n=x)$. The equation above translate to the following equation:

\begin{equation}
f_n(x)=\frac{2}{3}f_{n-1}(x)+\frac{1}{3}( f_{n-1} \ast f_{n-1})(x),
\label{s3r2}
\end{equation}

\noindent where $(a \ast b)(x)$ is the convolution $\int_{-\infty}^\infty a(y)
b(x-y) \dd y$, and $f_n(x)=0$ for $x<0$. The overlap difference
between two consecutive generations in the limit of large $n$, should
not alter the qualitative behavior of the probability distribution. We
therefore assume that there is a $n$-independent distribution $f$ such
that 

\begin{equation}
f_n(x) = f(x/\mu^n)/\mu^n,
\label{scalingrelation}
\end{equation}

\noindent for a scaling variable $\mu$. This is verified by numerical calculations, with $\mu=\langle Y_1\rangle=r^2/s
= 4/3$ which is the expected overlap at the first generation. $f(x)$ should then obey the equation

\begin{equation}
 f(x/\mu) /\mu=\frac{2}{3}f(x)+\frac{1}{3}( f \ast f)(x).
\label{31dist}
\end{equation}

To proceed with this functional equation we can take the Fourier
transform on both sides. Using the elementary properties of the
Fourier transformation we get

\begin{equation*}
\tilde f(k \mu ) =\frac{2}{3} \tilde f(k)+\frac{1}{3}\tilde f(k)^2.
\end{equation*}

If we assume that $\tilde f(k)$ is analytical in the complex plane, we can expand it as
$\tilde f(k) = \sum a_j k^j$. By matching terms we get a family of solutions

\begin{equation*}
a_0=1,\,\,\, a_1\in\mathbb{C},\,\,\, a_n=\frac{1}{ 4
  \left(\mu^{n-1}-1\right) }\sum_{j=1}^{n-1}a_ja_{n-j},\, n>2.
\end{equation*}

\noindent The real part of $\tilde f(k)$ has to be even, and the
imaginary part has to be odd to make $f(x)$ real. We therefore choose
$a_1$ to be purely imaginary.

Note that the Fourier transform is identical to the characteristic
function $\varphi(k)=\langle \exp(ikx) \rangle$ of
$f(x)$. The moments of the distribution is given by the relation
$\langle x^m \rangle = (-i)^m\varphi^{(m)}(0)$. The first moment of
$f(x)$ is by construction one, $\langle x \rangle = 1$, so we choose
$a_1=i$. ($a_0=1$ is consistent with the
normalization of $f(x)$.) All other moments are specified by
\begin{equation}
\langle x^m \rangle =(-i)^m\, m!\, a_m ,
\label{moments}
\end{equation}
\noindent and the probability distribution $f(x)$, is uniquely
determined in terms of its moments.

For the general random Cantor set, defined in the previous section, we
have the following results. For the overlap probability for the first
generation we have

\begin{equation*}
\Pr(Y_{1}^{(s,r)}= x) = p^{(s,r)}(x) = \frac{{r\choose x} {s-r \choose
    r-x}}{ {s \choose r}}.
\end{equation*}

\noindent Analogous to the distribution in eqn. (\ref{s3r2}), we have
that the overlap distribution for the general case should be of the form

\begin{equation}
f_{n}^{(s,r)}(x)=   \sum_{j=1}^r    p^{(s,r)}(j)    \left( 
  \underbrace{ f_{n-1}^{(s,r)}\ast \dots \ast
  f_{n-1}^{(s,r)}  }_{j-1 \text{ times}} \right) (x), 
\label{gendist1}
\end{equation}

\noindent and the $n$ invariant distribution $f^{(s,r)}(x)$ is given by

\begin{equation}
f^{(s,r)}(x/\mu)/\mu =   \sum_{j=1}^r    p^{(s,r)}(j)    \left( 
  \underbrace{ f^{(s,r)} \ast \dots \ast
  f^{(s,r)}  }_{j-1 \text{ times}} \right) (x),
\label{gendist}
\end{equation}

\noindent with $\mu=r^2/s$. 
By expanding the Fourier transform of $f^{(s,r)}(x)$ as an analytical function, we can specify
  the moments for the probability distribution as we did in eqn. (\ref{moments}), but the values of $s$ and $r$ have to be specified
  to do so.

\begin{figure}
  \subfloat[][]{\includegraphics[scale=0.6]{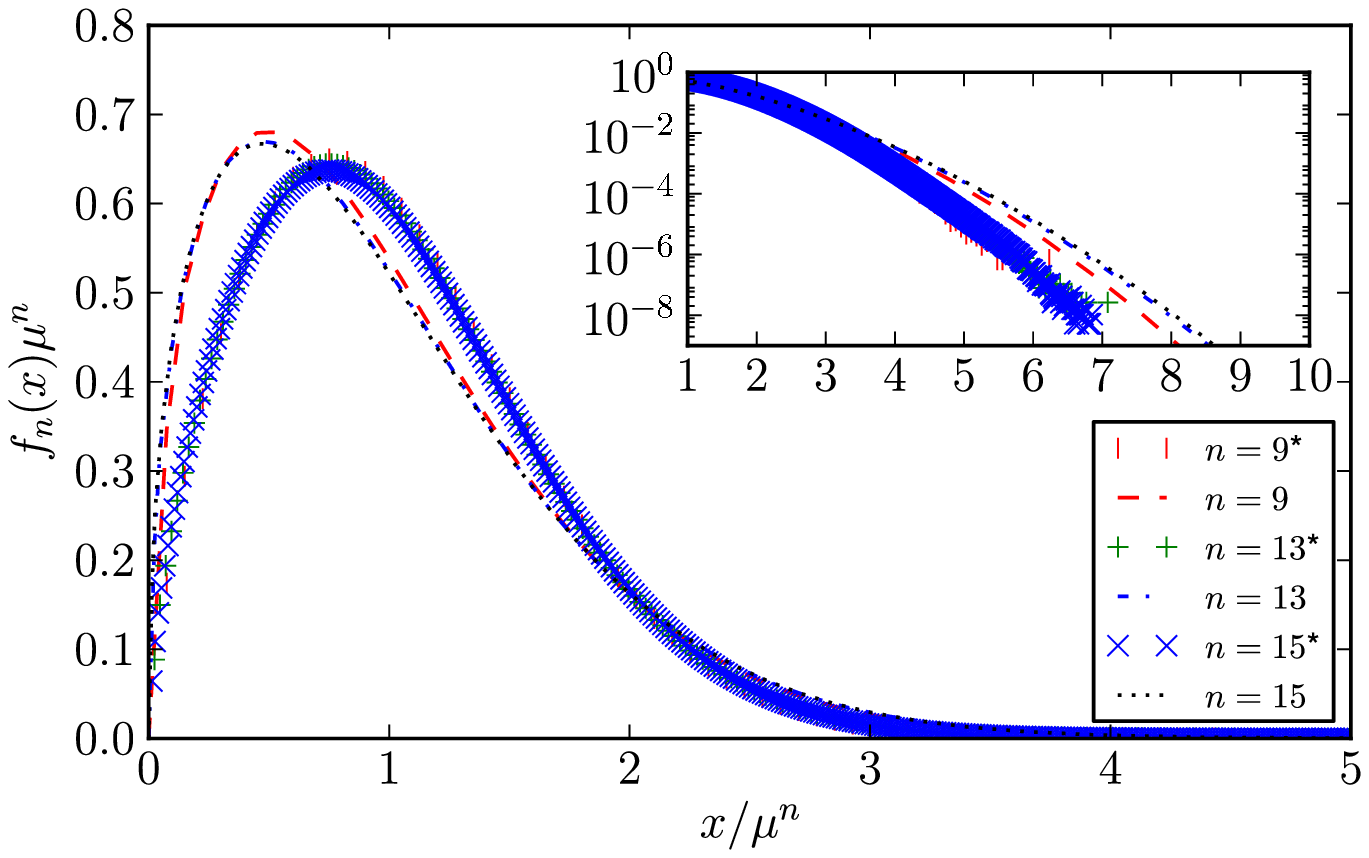}}\\
  \subfloat[][]{\includegraphics[scale=0.6]{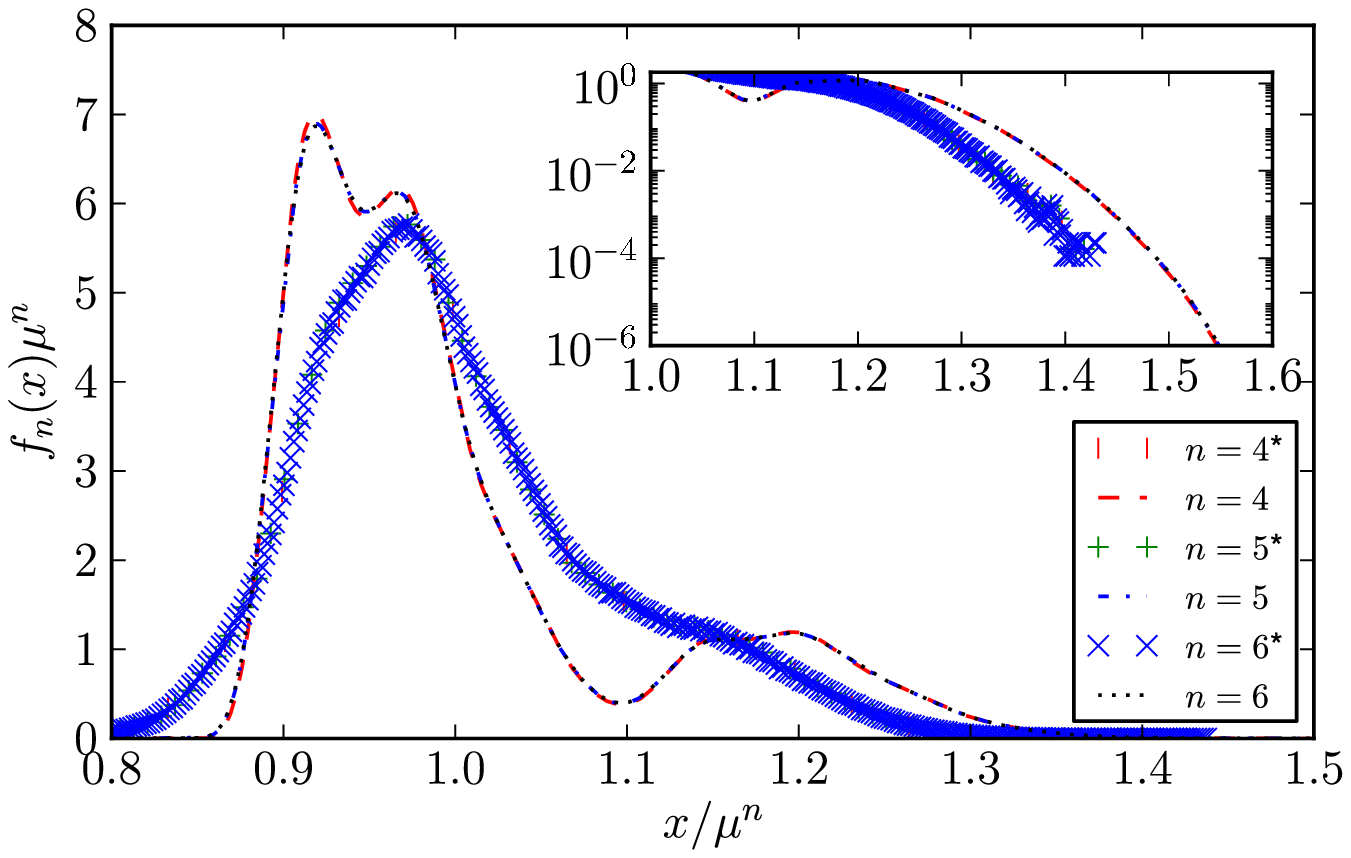}}
\caption{(Color online) The collapse of the overlap distribution for
    different generations ($n$ values). (a) for $s,r=3,2$ and (b) for $s,r=6,5$. The overlap distribution marked with
    lines are for random cantor sets, and are generated by eqn. (\ref{gendist1}).
   The overlap distribution marked with points (marked by an asterisk
   * in the legend), are for random cantor sets with a randomized offset, and are generated by sampling.  The
    $x$ axis is in units of expected overlap. The embedded plot shows the
    same in log scale for the $y$-axis.}
\label{plot1}
\end{figure}

\begin{figure}
  \begin{center}
 \subfloat[][]{\includegraphics[scale=0.6]{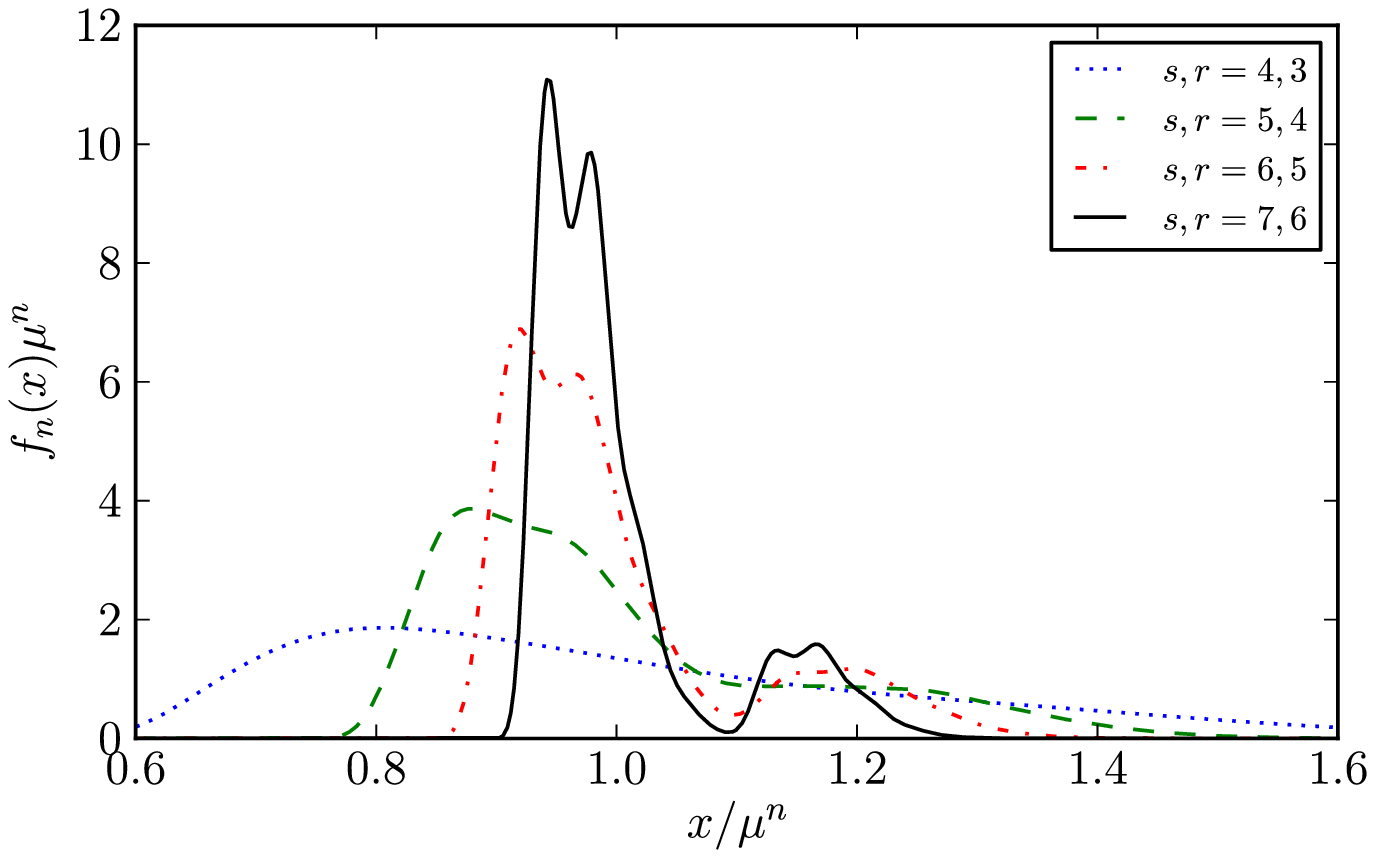}}\\
  \subfloat[][]{\includegraphics[scale=0.6]{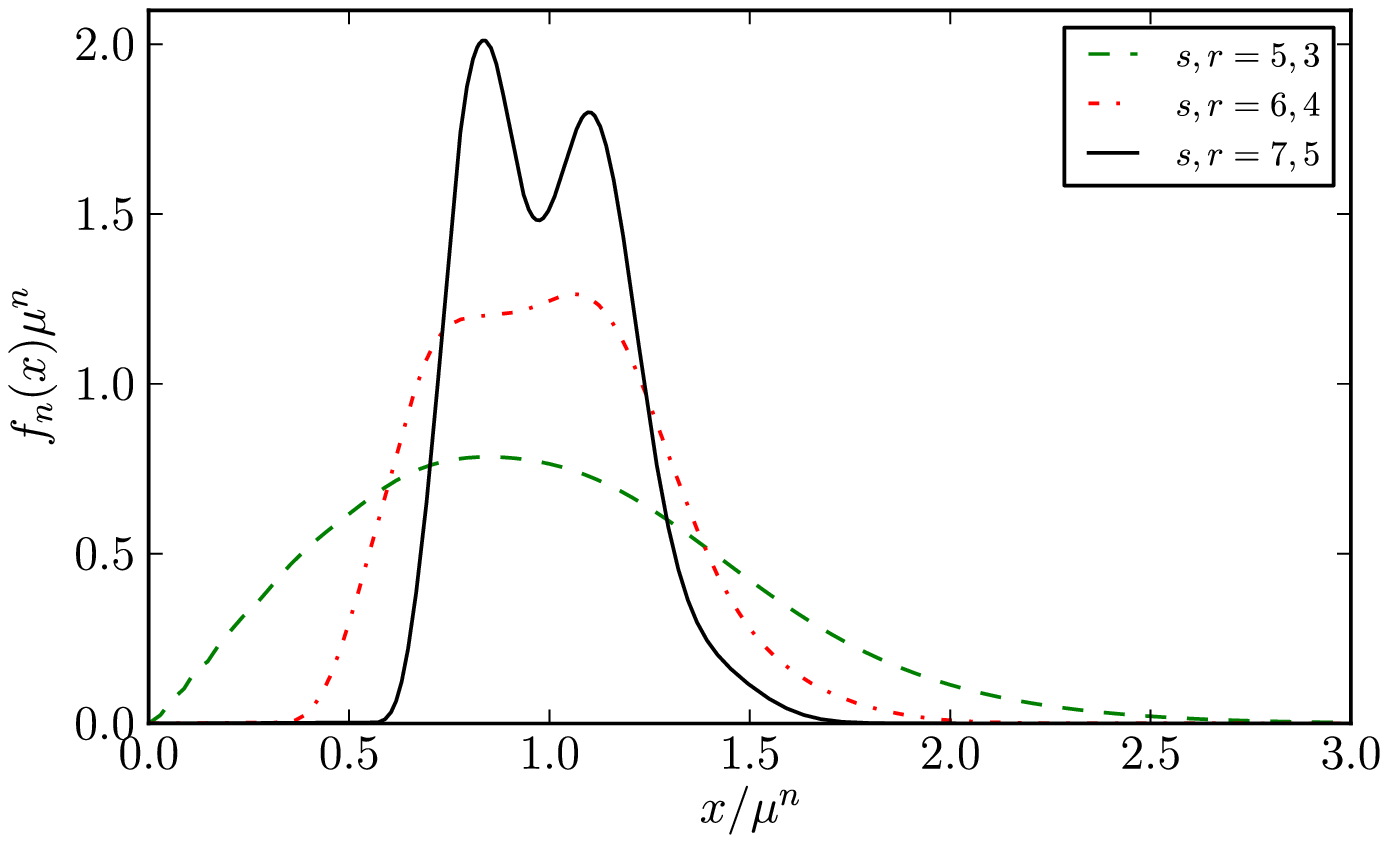}}
  \end{center}
  \caption{(Color online) The $n$ invariant overlap distribution,
    generated by eqn. (\ref{gendist1}), for
    different values of $s,r$, with  (a) $s-r=1$, and (b) $s-r=2$.  the
    $x$ axis is in units of expected overlap.}
\label{plot2}
\end{figure}

Up until now we have considered the overlap of the random Cantor sets
without any offset, i.e. the Cantor sets are placed such that the
first element of the top Cantor set (vector element $C_{n,1}$) overlap the first element of the lower
Cantor set (see Fig. \ref{model1} (a)). In order to study the overlap when
we also have a random offset, we assume
periodic boundary conditions as indicated by Fig. \ref{model1} (b). There is no obvious way to construct an
analytical expression for the distribution with the randomized offset, but we can
generate the distributions numerically. Fig. \ref{plot1} 
shows how the overlap distribution converges to the
$n$-invariant distribution with and without a randomized
offset, for (a) $s,r=3,2$, and (b) $s, r= 6,5$. 

\begin{figure}
 \subfloat[][]{\includegraphics[scale=0.6]{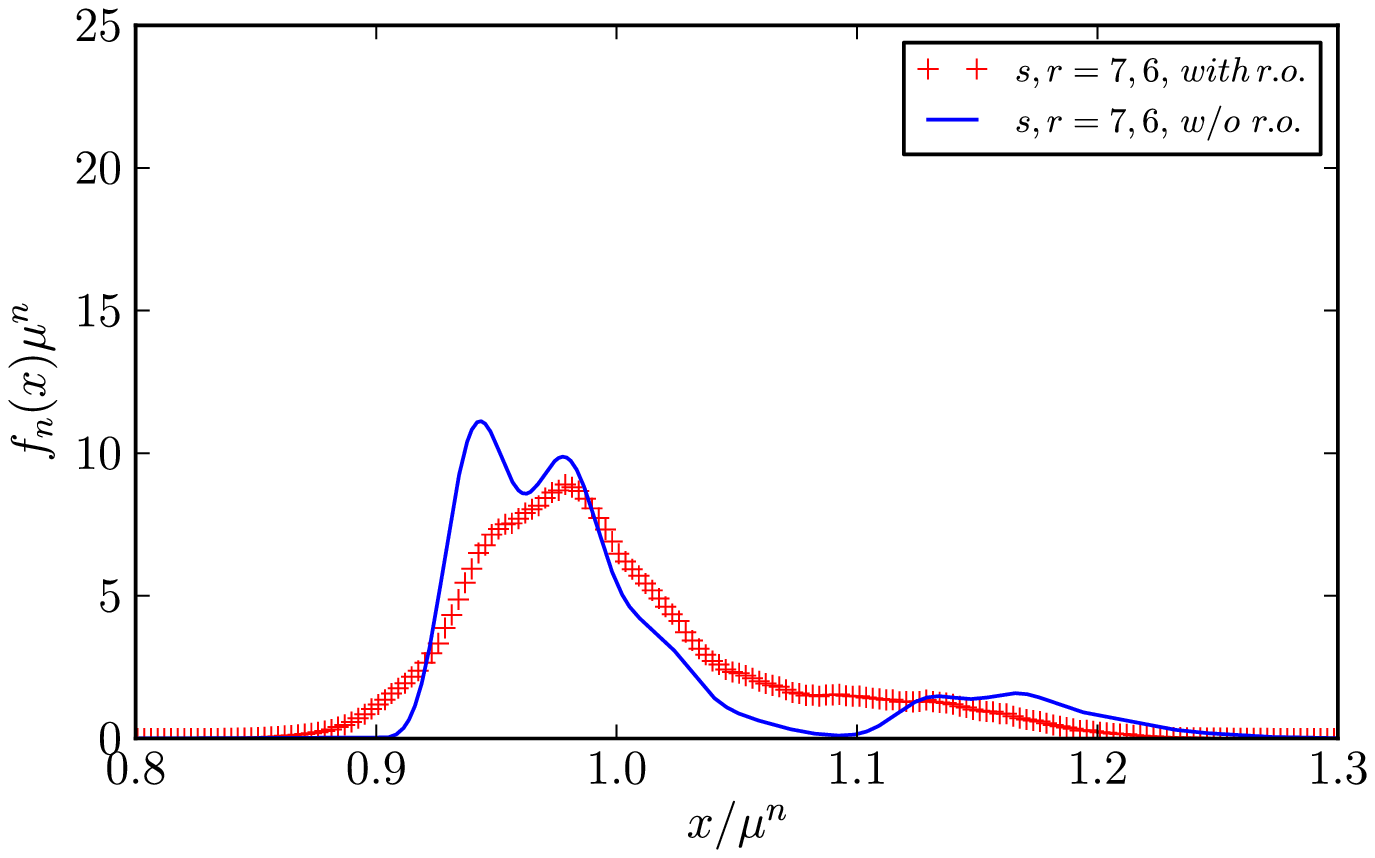}}\\
  \subfloat[][]{\includegraphics[scale=0.6]{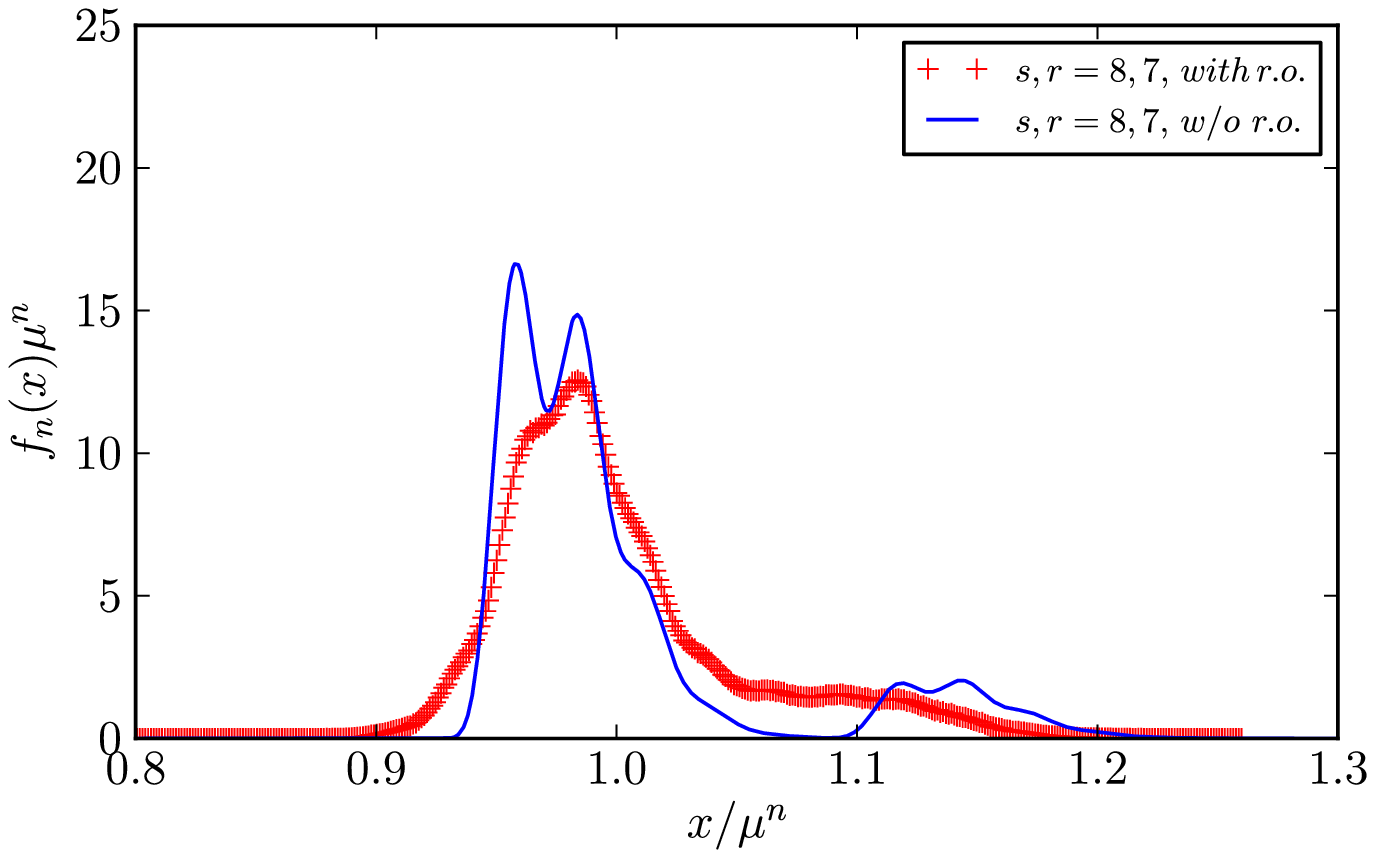}}\\
  \subfloat[][]{\includegraphics[scale=0.6]{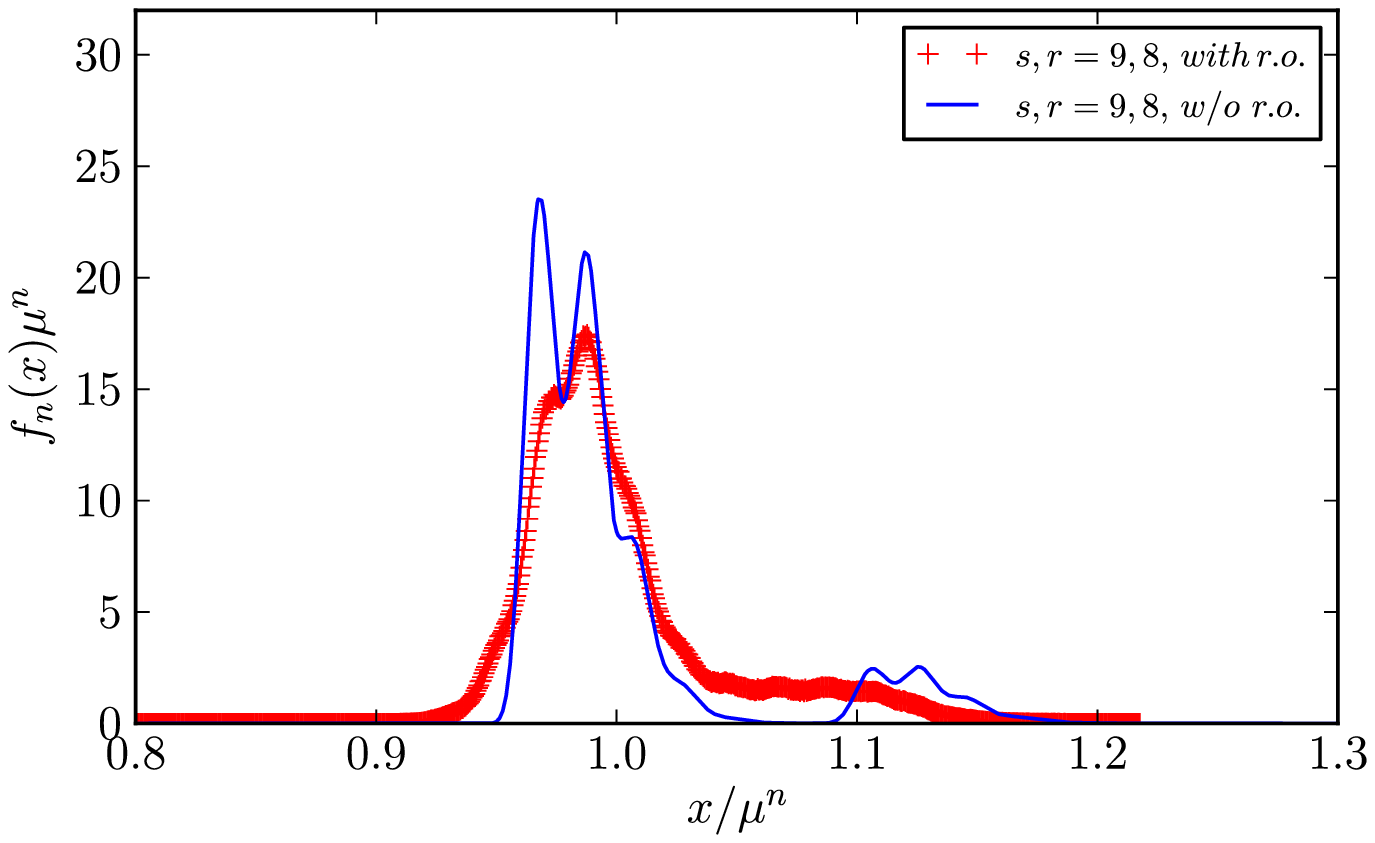}}
\caption{(Color online) The overlap distribution for $s-r=1$ when $r/s$
  goes to 1. $s, r = 7, 6$ in (a), $8, 7$ in (b) and $9, 8$ in
  (c). Lines (w/o randomized offset) are generated by
eqn. (\ref{gendist1}). Points (with randomized offset) are generated
by sampling. The $x$ axis is in units of expected overlap. }
\label{plot3}
\end{figure}

Fig. \ref{plot2} (a) shows the distributions for different
values of $s$ with $r=s-1$, and Fig. \ref{plot2} (b) shows the same for
$r=s-2$. The distributions are generated by
eqn. (\ref{gendist1}). Fig. \ref{plot3} shows how the overlap
distributions behave for $s-r=1$ when the ratio $r/s$ goes to one. The
overlap distribution with a randomized offset is generated by
sampling over the overlap for
1000 different configurations and for all the possible offsets.

\section{Summary and Discussion}
\noindent 
We model the static friction force between two atomically smooth surfaces
by a two-chain version of Tomlinson's model. We consider substitutional defects in
terms of vacancies along both chains (see Fig. \ref{model}). The
static friction force is assumed to be directly proportional to the
number of beads in the upper chain which are locked in the potential
wells in the potential arising from the chain beneath. This number is modeled by
the overlap of two binary chains with a given correlation structure.
We are in particular interested in the strong disorder limit where
self-similarity may appear. The self-similarity translates to the
power-law correlation in the spatial displacements of the remaining
beads. The main motivation to consider this is the
fact that the height profile of rough surfaces often
has a fractal (self-similar) property (see e.g. \cite{mandelbrot}).

A generalized random Cantor set is utilized
to capture such properties. This Cantor set is generated by removing
all but $r$ out of $s$ segments from each remaining element at each
generation. This procedure is explained in
detail in section \ref{cantorsec}. The remaining beads in the chain
are distributed according to this random Cantor set. We study both the cases
of with and without a random relative offset between the Cantor sets
(as illustrated in Fig. \ref{model1}).
The overlap of two chains is assumed to give the static friction
force. Hence, the distribution of
the static friction force is given by the distribution of the overlap
of the Cantor sets.

It may be noted here that the earlier applications of two fractal
overlap models \cite{chakrabarti,bhattacharyya,pathikrit1}, in the
context of earthquake dynamics, focused on the time series of the overlap of \emph{regular} (non-random)
Cantor sets. Our interest here is the overlap of \emph{random} Cantor sets in the context of static friction.

Cantor sets have already been used to represent the
 scale invariance property of the contact area overlap
 between two plastic surfaces 
 of macroscopic objects in the study by Warren and
 Krajcinovic \cite{warren}. Their model does not represent similar
 properties of each of the surfaces, as in our model. In the model presented here, Cantor sets (embedded
 in one dimension) are used to represent each of the
 surface and we calculate the overlap
 profile between them. Warren and Krajcinovic, on the other hand, use
 Cantor sets embedded in two dimensions, and calculate how it overlaps with a plane surface.
 As such, our model is different and the results are not comparable.

For the overlap of the generalized random Cantor sets, 
without the randomized offset, we have found a recurrence relation
for the distribution (eqn. (\ref{gendist1})). Moreover we find that
this distribution follows a scaling structure
(eqn. (\ref{scalingrelation})). This scaling leads to a distribution which is
independent of the generation $n$ (though dependent on values of $s$
and $r$). We further show how one can specify the
  distribution uniquely in terms of the moments, and do the
  calculation explicitly for the case when
  $s=3$ and $r=2$ (see eqn. (\ref{moments})).

After introducing a randomized offset between the Cantor sets (with periodic
boundary conditions), we no longer have a
recurrence relation, but we find numerically similar qualitative
behavior. The distribution for the case with a randomized offset shows
the same $n$ independent scaling behavior as the distribution without a randomized
offset. Moreover, the distribution is shifted to slightly higher values of overlap as shown in Fig. \ref{plot1}. The same behavior is seen for different values of $s$ and
$r$. The embedded log plot in Fig. \ref{plot1} shows that the tail
behavior falls faster than exponential in both cases.

When we look at the distribution for different values of $s$ and $r$,
generated by eqn. (\ref{gendist1}), we see the emergence of multiple
local maxima.  This is a property of the sum of convolutions coming
from eqn. (\ref{gendist1}), and depends on the allowed values of overlap for the
first generation. These local maxima are averaged out when we
look at the case with a randomized offset (see Fig. \ref{plot1} and
Fig. \ref{plot3}). 
We have conveniently presented the overlap distributions in units of
the expected overlap (the actual overlap of a given set is found by
multiplying the $x$ axis with $\mu^n = (r^2/s)^n$).

For a realistic chain for beads with vacancies, we can not assume that
the remaining beads are distributed according to precise values of $s$
and $r$. Nor can we assume that the overlap distribution is as for two Cantor sets
without randomizing a relative offset. If we consider the
limit where $r/s$ approaches unity, (i.e. the
limit where only a small fraction of beads are removed at every
generation), we find for the overlap distribution with a randomized offset, that the shape of the distributions have some common general
properties. The distribution gets a peak at the expected overlap, but
also an interval with a non-zero probability for values higher than
the expected overlap, see Fig. \ref{plot3}.

The overlap of beads distributed as a Cantor set is qualitatively very different
from the overlap of uniformly displaced beads. To compare the distribution with that in
Fig. \ref{plot2}, set
$p=(r/s)^n$ and $N=s^n$ in eqn. (\ref{randoverlap}). The resulting
distribution would then be approximated by a Gaussian in the limit of large $n$, with
mean $N p^2=(r^2/s)^n$ and a variance $Np^2(1-p^2)= (r^2/s)^n(1-(r^2/s)^n)$. This would
correspond to a single peak at $x/\mu^n=1$ in
Figs. \ref{plot1}-\ref{plot3}. On the contrary the static friction
force distribution for chain models, where the remaining beads are distributed in
a scale invariant way, will not converge to a delta peak distribution
but rather would be like that for Cantor sets with a randomized offset
as in Figs. \ref{plot1}-\ref{plot3}.

It is hard to compare our results with experimental data on
microscopic dry friction between surfaces having scale invariant
disorder. Surfaces having microscopic self-affine
disorder have been studied using
 AFM \cite{fractalafm}, but we are unfortunately not aware of any study where the
static friction force between two such surfaces have been considered.
 We would like to mention that the above results for the distribution microscopic friction
qualitatively agrees with the observation of similar distributions
of the friction coefficient of Aluminium  alloys under cold
rolling (see e.g., \cite{liu}). Such an analysis can also be
effectively utilized for comparing the distributions for
dry friction coefficients between plastic rock surfaces having
well known scaling properties  of asperities. However, in
a different context of studying the effect of multiscale roughness
on contact mechanics, similar analysis has already been done (see e.g., \cite{sokoloff}).

\bibliography{bibliography}

\begin{thebibliography}{10}%
\makeatletter
\providecommand \@ifxundefined [1]{%
 \ifx #1\undefined \expandafter \@firstoftwo
 \else \expandafter \@secondoftwo
\fi
}%
\providecommand \@ifnum [1]{%
 \ifnum #1\expandafter \@firstoftwo
 \else \expandafter \@secondoftwo
\fi
}%
\providecommand \enquote [1]{``#1''}%
\providecommand \bibnamefont  [1]{#1}%
\providecommand \bibfnamefont [1]{#1}%
\providecommand \citenamefont [1]{#1}%
\providecommand\href[0]{\@sanitize\@href}%
\providecommand\@href[1]{\endgroup\@@startlink{#1}\endgroup\@@href}%
\providecommand\@@href[1]{#1\@@endlink}%
\providecommand \@sanitize [0]{\begingroup\catcode`\&12\catcode`\#12\relax}%
\@ifxundefined \pdfoutput {\@firstoftwo}{%
 \@ifnum{\z@=\pdfoutput}{\@firstoftwo}{\@secondoftwo}%
}{%
 \providecommand\@@startlink[1]{\leavevmode\special{html:<a href="#1">}}%
 \providecommand\@@endlink[0]{\special{html:</a>}}%
}{%
 \providecommand\@@startlink[1]{%
  \leavevmode
  \pdfstartlink
   attr{/Border[0 0 1 ]/H/I/C[0 1 1]}%
   user{/Subtype/Link/A<</Type/Action/S/URI/URI(#1)>>}%
  \relax
 }%
 \providecommand\@@endlink[0]{\pdfendlink}%
}%
\providecommand \url  [0]{\begingroup\@sanitize \@url }%
\providecommand \@url [1]{\endgroup\@href {#1}{\urlprefix}}%
\providecommand \urlprefix [0]{URL }%
\providecommand \Eprint[0]{\href }%
\@ifxundefined \urlstyle {%
  \providecommand \doi [1]{doi:\discretionary{}{}{}#1}%
}{%
  \providecommand \doi [0]{doi:\discretionary{}{}{}\begingroup
  \urlstyle{rm}\Url }%
}%
\providecommand \doibase [0]{http://dx.doi.org/}%
\providecommand \Doi[1]{\href{\doibase#1}}%
\providecommand \bibAnnote [3]{%
  \BibitemShut{#1}%
  \begin{quotation}\noindent
    \textsc{Key:}\ #2\\\textsc{Annotation:}\ #3%
  \end{quotation}%
}%
\providecommand \bibAnnoteFile [2]{%
  \IfFileExists{#2}{\bibAnnote {#1} {#2} {\input{#2}}}{}%
}%
\providecommand \typeout [0]{\immediate \write \m@ne }%
\providecommand \selectlanguage [0]{\@gobble}%
\providecommand \bibinfo [0]{\@secondoftwo}%
\providecommand \bibfield [0]{\@secondoftwo}%
\providecommand \translation [1]{[#1]}%
\providecommand \BibitemOpen[0]{}%
\providecommand \bibitemStop [0]{}%
\providecommand \bibitemNoStop [0]{.\EOS\space}%
\providecommand \EOS [0]{\spacefactor3000\relax}%
\providecommand \BibitemShut [1]{\csname bibitem#1\endcsname}%
\bibitem{bowden}%
  \BibitemOpen
  \bibfield{author}{%
  \bibinfo {author} {\bibfnamefont{F.~P.}\ \bibnamefont{Bowden}}\ and\ \bibinfo
  {author} {\bibfnamefont{D.}~\bibnamefont{Tabor}},\ }%
  \emph{\bibinfo {title} {The friction and Lubrication of Solids}}\ (\bibinfo
  {publisher} {Clarendon Press, Oxford},\ \bibinfo {year} {1954})%
  \bibAnnoteFile{NoStop}{bowden}%
\bibitem{bnj}%
  \BibitemOpen
  \bibfield{author}{%
  \bibinfo {author} {\bibfnamefont{B.~N.~J.}\ \bibnamefont{Persson}},\ }%
  \emph{\bibinfo {title} {Sliding Friction: Physical Principles and
  Application}}\ (\bibinfo {publisher} {2nd ed., Springer, Berlin},\ \bibinfo
  {year} {2000})%
  \bibAnnoteFile{NoStop}{bnj}%
\bibitem{braun}%
  \BibitemOpen
  \bibfield{author}{%
  \bibinfo {author} {\bibfnamefont{O.~M.}\ \bibnamefont{Baun}}\ and\ \bibinfo
  {author} {\bibfnamefont{A.~G.}\ \bibnamefont{Naumovets}},\ }%
  \bibfield{journal}{%
  \bibinfo {journal} {Surf. Sc. Rep.}\ }%
  \textbf{\bibinfo {volume} {60}},\ \bibinfo {pages} {79} (\bibinfo {year}
  {2006})%
  \bibAnnoteFile{NoStop}{braun}%
\bibitem{gnecco}%
  \BibitemOpen
  \bibfield{author}{%
  \bibinfo {author} {\bibfnamefont{E.}~\bibnamefont{Gnecco}}, \bibinfo {author}
  {\bibfnamefont{R.}~\bibnamefont{Bennewitz}}, \bibinfo {author}
  {\bibfnamefont{T.}~\bibnamefont{Gyalog}}, \bibinfo {author}
  {\bibfnamefont{C.}~\bibnamefont{Loppacher}}, \bibinfo {author}
  {\bibfnamefont{M.}~\bibnamefont{Bammerlin}}, \bibinfo {author}
  {\bibfnamefont{E.}~\bibnamefont{Meyer}},\ and\ \bibinfo {author}
  {\bibfnamefont{H.~J.}\ \bibnamefont{Guntherodt}},\ }%
  \bibfield{journal}{%
  \bibinfo {journal} {Phys. Rev. Lett.}\ }%
  \textbf{\bibinfo {volume} {84}},\ \bibinfo {pages} {1172} (\bibinfo {year}
  {2000})%
  \bibAnnoteFile{NoStop}{gnecco}%
\bibitem{sang}%
  \BibitemOpen
  \bibfield{author}{%
  \bibinfo {author} {\bibfnamefont{Y.}~\bibnamefont{Sang}}, \bibinfo {author}
  {\bibfnamefont{M.}~\bibnamefont{Dube}},\ and\ \bibinfo {author}
  {\bibfnamefont{M.}~\bibnamefont{Grant}},\ }%
  \bibfield{journal}{%
  \bibinfo {journal} {Phys. Rev. lett.}\ }%
  \textbf{\bibinfo {volume} {87}},\ \bibinfo {pages} {174301} (\bibinfo {year}
  {2001})%
  \bibAnnoteFile{NoStop}{sang}%
\bibitem{kajita}%
  \BibitemOpen
  \bibfield{author}{%
  \bibinfo {author} {\bibfnamefont{S.}~\bibnamefont{Kajita}}, \bibinfo {author}
  {\bibfnamefont{H.}~\bibnamefont{Washizu}},\ and\ \bibinfo {author}
  {\bibfnamefont{T.}~\bibnamefont{Ohomori}},\ }%
  \bibfield{journal}{%
  \bibinfo {journal} {Europhys. Lett.}\ }%
  \textbf{\bibinfo {volume} {87}},\ \bibinfo {pages} {66002} (\bibinfo {year}
  {2009})%
  \bibAnnoteFile{NoStop}{kajita}%
\bibitem{capozza}%
  \BibitemOpen
  \bibfield{author}{%
  \bibinfo {author} {\bibfnamefont{R.}~\bibnamefont{Capozza}}, \bibinfo
  {author} {\bibfnamefont{A.}~\bibnamefont{Vanossi}}, \bibinfo {author}
  {\bibfnamefont{A.}~\bibnamefont{Vezzani}},\ and\ \bibinfo {author}
  {\bibfnamefont{S.}~\bibnamefont{Zapperi}},\ }%
  \bibfield{journal}{%
  \bibinfo {journal} {Phys. Rev. Lett.}\ }%
  \textbf{\bibinfo {volume} {103}},\ \bibinfo {pages} {085502} (\bibinfo {year}
  {2009})%
  \bibAnnoteFile{NoStop}{capozza}%
\bibitem{kawaguchi}%
  \BibitemOpen
  \bibfield{author}{%
  \bibinfo {author} {\bibfnamefont{T.}~\bibnamefont{Kawaguchi}}\ and\ \bibinfo
  {author} {\bibfnamefont{H.}~\bibnamefont{Matsukawa}},\ }%
  \bibfield{journal}{%
  \bibinfo {journal} {Phys. Rev. B}\ }%
  \textbf{\bibinfo {volume} {56}},\ \bibinfo {pages} {4261} (\bibinfo {year}
  {1997})%
  \bibAnnoteFile{NoStop}{kawaguchi}%
\bibitem{caroli2}%
  \BibitemOpen
  \bibfield{author}{%
  \bibinfo {author} {\bibfnamefont{F.}~\bibnamefont{Heslot}}, \bibinfo {author}
  {\bibfnamefont{T.}~\bibnamefont{Baumberger}}, \bibinfo {author}
  {\bibfnamefont{B.}~\bibnamefont{Perrin}}, \bibinfo {author}
  {\bibfnamefont{B.}~\bibnamefont{Caroli}},\ and\ \bibinfo {author}
  {\bibfnamefont{C.}~\bibnamefont{Caroli}},\ }%
  \bibfield{journal}{%
  \bibinfo {journal} {Phys. Rev. E}\ }%
  \textbf{\bibinfo {volume} {49}},\ \bibinfo {pages} {4973} (\bibinfo {year}
  {1994})%
  \bibAnnoteFile{NoStop}{caroli2}%
\bibitem{mate}%
  \BibitemOpen
  \bibfield{author}{%
  \bibinfo {author} {\bibfnamefont{C.~M.}\ \bibnamefont{Mate}}, \bibinfo
  {author} {\bibfnamefont{G.}~\bibnamefont{McClelland}}, \bibinfo {author}
  {\bibfnamefont{R.}~\bibnamefont{Erlandsson}},\ and\ \bibinfo {author}
  {\bibfnamefont{S.}~\bibnamefont{Chiang}},\ }%
  \bibfield{journal}{%
  \bibinfo {journal} {Phys. Rev. Lett.}\ }%
  \textbf{\bibinfo {volume} {59}},\ \bibinfo {pages} {1942} (\bibinfo {year}
  {1987})%
  \bibAnnoteFile{NoStop}{mate}%
\bibitem{hirano}%
  \BibitemOpen
  \bibfield{author}{%
  \bibinfo {author} {\bibfnamefont{M.}~\bibnamefont{Hirano}}, \bibinfo {author}
  {\bibfnamefont{K.}~\bibnamefont{Shinjo}}, \bibinfo {author}
  {\bibfnamefont{R.}~\bibnamefont{Kaneko}},\ and\ \bibinfo {author}
  {\bibfnamefont{Y.}~\bibnamefont{Murata}},\ }%
  \bibfield{journal}{%
  \bibinfo {journal} {Phys. Rev. Lett.}\ }%
  \textbf{\bibinfo {volume} {78}},\ \bibinfo {pages} {1448} (\bibinfo {year}
  {1997})%
  \bibAnnoteFile{NoStop}{hirano}%
\bibitem{caroli}%
  \BibitemOpen
  \bibfield{author}{%
  \bibinfo {author} {\bibfnamefont{T.}~\bibnamefont{Baumberger}}, \bibinfo
  {author} {\bibfnamefont{P.}~\bibnamefont{Berthoud}},\ and\ \bibinfo {author}
  {\bibfnamefont{C.}~\bibnamefont{Caroli}},\ }%
  \bibfield{journal}{%
  \bibinfo {journal} {Phys. Rev. B}\ }%
  \textbf{\bibinfo {volume} {60}},\ \bibinfo {pages} {3928} (\bibinfo {year}
  {1999})%
  \bibAnnoteFile{NoStop}{caroli}%
\bibitem{muser}%
  \BibitemOpen
  \bibfield{author}{%
  \bibinfo {author} {\bibfnamefont{G.}~\bibnamefont{He}}, \bibinfo {author}
  {\bibfnamefont{M.}~\bibnamefont{Muser}},\ and\ \bibinfo {author}
  {\bibfnamefont{M.}~\bibnamefont{Robbins}},\ }%
  \bibfield{journal}{%
  \bibinfo {journal} {Science}\ }%
  \textbf{\bibinfo {volume} {284}},\ \bibinfo {pages} {5420} (\bibinfo {year}
  {1999})%
  \bibAnnoteFile{NoStop}{muser}%
\bibitem{tomlinson}%
  \BibitemOpen
  \bibfield{author}{%
  \bibinfo {author} {\bibfnamefont{G.~A.}\ \bibnamefont{Tomlinson}},\ }%
  \bibfield{journal}{%
  \bibinfo {journal} {Philos. Mag.}\ }%
  \textbf{\bibinfo {volume} {7}},\ \bibinfo {pages} {905} (\bibinfo {year}
  {1929})%
  \bibAnnoteFile{NoStop}{tomlinson}%
\bibitem{weiss}%
  \BibitemOpen
  \bibfield{author}{%
  \bibinfo {author} {\bibfnamefont{M.}~\bibnamefont{Weiss}}\ and\ \bibinfo
  {author} {\bibfnamefont{F.~J.}\ \bibnamefont{Elmer}},\ }%
  \bibfield{journal}{%
  \bibinfo {journal} {Phys. Rev. B}\ }%
  \textbf{\bibinfo {volume} {53}},\ \bibinfo {pages} {7539} (\bibinfo {year}
  {1996})%
  \bibAnnoteFile{NoStop}{weiss}%
\bibitem{fk1}%
  \BibitemOpen
  \bibfield{author}{%
  \bibinfo {author} {\bibfnamefont{Y.}~\bibnamefont{Frenkel}}\ and\ \bibinfo
  {author} {\bibfnamefont{T.}~\bibnamefont{Kontorova}},\ }%
  \bibfield{journal}{%
  \bibinfo {journal} {Zh. Eksp. Teor. Phys.}\ }%
  \textbf{\bibinfo {volume} {8}},\ \bibinfo {pages} {1340} (\bibinfo {year}
  {1938})%
  \bibAnnoteFile{NoStop}{fk1}%
\bibitem{mandelbrot}%
  \BibitemOpen
  \bibfield{author}{%
  \bibinfo {author} {\bibfnamefont{B.~B.}\ \bibnamefont{Mandelbrot}},\ }%
  \emph{\bibinfo {title} {The Fractal Geometry of nature}}\ (\bibinfo
  {publisher} {Freeman, San Francisco},\ \bibinfo {year} {1982})%
  \bibAnnoteFile{NoStop}{mandelbrot}%
\bibitem{falconer}%
  \BibitemOpen
  \bibfield{author}{%
  \bibinfo {author} {\bibfnamefont{K.}~\bibnamefont{Falconer}},\ }%
  \emph{\bibinfo {title} {Fractal Geometry and its Applications}}\ (\bibinfo
  {publisher} {Wiley, Chichester},\ \bibinfo {year} {1999})%
  \bibAnnoteFile{NoStop}{falconer}%
\bibitem{chakrabarti}%
  \BibitemOpen
  \bibfield{author}{%
  \bibinfo {author} {\bibfnamefont{B.~K.}\ \bibnamefont{Chakrabarti}}\ and\
  \bibinfo {author} {\bibfnamefont{R.~B.}\ \bibnamefont{Stinchcombe}},\ }%
  \bibfield{journal}{%
  \bibinfo {journal} {Physica A}\ }%
  \textbf{\bibinfo {volume} {270}},\ \bibinfo {pages} {27} (\bibinfo {year}
  {1999})%
  \bibAnnoteFile{NoStop}{chakrabarti}%
\bibitem{bhattacharyya}%
  \BibitemOpen
  \bibfield{author}{%
  \bibinfo {author} {\bibfnamefont{P.}~\bibnamefont{Bhattacharyya}},\ }%
  \bibfield{journal}{%
  \bibinfo {journal} {Physica A}\ }%
  \textbf{\bibinfo {volume} {348}},\ \bibinfo {pages} {199} (\bibinfo {year}
  {2005})%
  \bibAnnoteFile{NoStop}{bhattacharyya}%
\bibitem{pathikrit1}%
  \BibitemOpen
  \bibfield{author}{%
  \bibinfo {author} {\bibfnamefont{P.}~\bibnamefont{Bhattacharya}}, \bibinfo
  {author} {\bibfnamefont{B.~K.}\ \bibnamefont{Chakrabarti}}, \bibinfo {author}
  {\bibnamefont{Kamal}},\ and\ \bibinfo {author}
  {\bibfnamefont{D.}~\bibnamefont{Samanta}},\ }%
  in\ \emph{\bibinfo {booktitle} {Reviews of Nonlinear Dynamics and
  Complexity}},\ \bibinfo {editor} {edited by\ \bibinfo {editor}
  {\bibfnamefont{H.~G.}\ \bibnamefont{Schuster}}}\ (\bibinfo {publisher} {Wiley
  - VCH, Weinheim},\ \bibinfo {year} {2009})\ pp.\ \bibinfo {pages} {107--158}%
  \bibAnnoteFile{NoStop}{pathikrit1}%
\bibitem{warren}%
  \BibitemOpen
  \bibfield{author}{%
  \bibinfo {author} {\bibfnamefont{T.~L.}\ \bibnamefont{Warren}}\ and\ \bibinfo
  {author} {\bibfnamefont{D.}~\bibnamefont{Krajcinovic}},\ }%
  \bibfield{journal}{%
  \bibinfo {journal} {Wear}\ }%
  \textbf{\bibinfo {volume} {196}},\ \bibinfo {pages} {1} (\bibinfo {year}
  {1996})%
  \bibAnnoteFile{NoStop}{warren}%
\bibitem{fractalafm}%
  \BibitemOpen
  \bibfield{author}{%
  \bibinfo {author} {\bibfnamefont{R.}~\bibnamefont{Buzio}}, \bibinfo {author}
  {\bibfnamefont{C.}~\bibnamefont{Boragno}},\ and\ \bibinfo {author}
  {\bibfnamefont{U.}~\bibnamefont{Valbusa}},\ }%
  \bibfield{journal}{%
  \bibinfo {journal} {Wear}\ }%
  \textbf{\bibinfo {volume} {254}},\ \bibinfo {pages} {917} (\bibinfo {year}
  {2003})%
  \bibAnnoteFile{NoStop}{fractalafm}%
\bibitem{liu}%
  \BibitemOpen
  \bibfield{author}{%
  \bibinfo {author} {\bibfnamefont{Y.~J.}\ \bibnamefont{Liu}}, \bibinfo
  {author} {\bibfnamefont{D.~D.}\ \bibnamefont{Tieu}, \bibfnamefont{A.~K.
  adn~Wang}},\ and\ \bibinfo {author} {\bibfnamefont{D.}~\bibnamefont{Yuen}},\
  }%
  \bibfield{journal}{%
  \bibinfo {journal} {J. Mat. Process. Tech.}\ }%
  \textbf{\bibinfo {volume} {111}},\ \bibinfo {pages} {142} (\bibinfo {year}
  {2001})%
  \bibAnnoteFile{NoStop}{liu}%
\bibitem{sokoloff}%
  \BibitemOpen
  \bibfield{author}{%
  \bibinfo {author} {\bibfnamefont{J.~B.}\ \bibnamefont{Sokoloff}},\ }%
  \bibfield{journal}{%
  \bibinfo {journal} {Phys. Rev. E.}\ }%
  \textbf{\bibinfo {volume} {78}},\ \bibinfo {pages} {036111} (\bibinfo {year}
  {2008})%
  \bibAnnoteFile{NoStop}{sokoloff}%
\end{thebibliography}%

\end{document}